\documentclass[12pt]{article}
\usepackage{graphicx}
\usepackage{lscape}
\usepackage{citesort}
\usepackage{amssymb}
\usepackage{appendix}
\usepackage{multirow}
\usepackage[table]{xcolor}
\usepackage{colortbl}
\definecolor{lightgray}{gray}{0.9}
\usepackage{mathrsfs}
\begin{document}
\def\qq{\langle \bar q q \rangle}
\def\uu{\langle \bar u u \rangle}
\def\dd{\langle \bar d d \rangle}
\def\sp{\langle \bar s s \rangle}
\def\GG{\langle g_s^2 G^2 \rangle}
\def\Tr{\mbox{Tr}}
\def\figt#1#2#3{
        \begin{figure}
        $\left. \right.$
        \vspace*{-2cm}
        \begin{center}
        \includegraphics[width=10cm]{#1}
        \end{center}
        \vspace*{-0.2cm}
        \caption{#3}
        \label{#2}
        \end{figure}
    }

\def\figb#1#2#3{
        \begin{figure}
        $\left. \right.$
        \vspace*{-1cm}
        \begin{center}
        \includegraphics[width=10cm]{#1}
        \end{center}
        \vspace*{-0.2cm}
        \caption{#3}
        \label{#2}
        \end{figure}
                }

\def\ds{\displaystyle}
\def\beq{\begin{equation}}
\def\eeq{\end{equation}}
\def\bea{\begin{eqnarray}}
\def\eea{\end{eqnarray}}
\def\beeq{\begin{eqnarray}}
\def\eeeq{\end{eqnarray}}
\def\ve{\vert}
\def\vel{\left|}
\def\ver{\right|}
\def\nnb{\nonumber}
\def\ga{\left(}
\def\dr{\right)}
\def\aga{\left\{}
\def\adr{\right\}}
\def\lla{\left<}
\def\rra{\right>}
\def\rar{\rightarrow}
\def\lrar{\leftrightarrow}
\def\nnb{\nonumber}
\def\la{\langle}
\def\ra{\rangle}
\def\ba{\begin{array}}
\def\ea{\end{array}}
\def\tr{\mbox{Tr}}
\def\ssp{{\Sigma^{*+}}}
\def\sso{{\Sigma^{*0}}}
\def\ssm{{\Sigma^{*-}}}
\def\xis0{{\Xi^{*0}}}
\def\xism{{\Xi^{*-}}}
\def\qs{\la \bar s s \ra}
\def\qu{\la \bar u u \ra}
\def\qd{\la \bar d d \ra}
\def\qq{\la \bar q q \ra}
\def\gGgG{\la g^2 G^2 \ra}
\def\q{\gamma_5 \not\!q}
\def\x{\gamma_5 \not\!x}
\def\g5{\gamma_5}
\def\sb{S_Q^{cf}}
\def\sd{S_d^{be}}
\def\su{S_u^{ad}}
\def\sbp{{S}_Q^{'cf}}
\def\sdp{{S}_d^{'be}}
\def\sup{{S}_u^{'ad}}
\def\ssp{{S}_s^{'??}}

\def\sig{\sigma_{\mu \nu} \gamma_5 p^\mu q^\nu}
\def\fo{f_0(\frac{s_0}{M^2})}
\def\ffi{f_1(\frac{s_0}{M^2})}
\def\fii{f_2(\frac{s_0}{M^2})}
\def\O{{\cal O}}
\def\sl{{\Sigma^0 \Lambda}}
\def\es{\!\!\! &=& \!\!\!}
\def\ap{\!\!\! &\approx& \!\!\!}
\def\md{\!\!\!\! &\mid& \!\!\!\!}
\def\ar{&+& \!\!\!}
\def\ek{&-& \!\!\!}
\def\kek{\!\!\!&-& \!\!\!}
\def\cp{&\times& \!\!\!}
\def\se{\!\!\! &\simeq& \!\!\!}
\def\eqv{&\equiv& \!\!\!}
\def\kpm{&\pm& \!\!\!}
\def\kmp{&\mp& \!\!\!}
\def\mcdot{\!\cdot\!}
\def\erar{&\rightarrow&}
\def\olra{\stackrel{\leftrightarrow}}
\def\ola{\stackrel{\leftarrow}}
\def\ora{\stackrel{\rightarrow}}

\def\simlt{\stackrel{<}{{}_\sim}}
\def\simgt{\stackrel{>}{{}_\sim}}


\title{
         {\Large
                 {\bf
                      Modifications on nucleon parameters at finite temperature 
                 }
         }
      }

\author{ K. Azizi \thanks {e-mail: kazizi@dogus.edu.tr}\,\,, G. Kaya \thanks {e-mail: gkaya@dogus.edu.tr}  \\
Department of Physics, Do\u gu\c s
University,
 Ac{\i}badem-Kad{\i}k\"oy, \\ 34722 Istanbul, Turkey\\
 }
\date{}

\begin{titlepage}
\maketitle
\thispagestyle{empty}

\begin{abstract}
Taking into account the additional operators coming up at finite temperature, we investigate the mass and residue of the nucleon 
in the framework of thermal QCD sum rules. We observe that the mass and residue
of the nucleon are initially  insusceptible to increase of temperature, however after a certain temperature, 
they start to fall increasing the temperature. 

\end{abstract}

~~~PACS number(s): 11.10.Wx, 11.55.Hx, 14.20.-c, 14.20.Dh
\end{titlepage}

\section{Introduction}

The investigation of the modifications on the hadronic parameters such as masses, decay constants, widths and coupling constants, etc. 
at finite temperature is one of the important research areas 
of thermal QCD. The theoretical determinations of these modifications provide us with the better understanding of 
the properties of hot and dense QCD matter produced by the heavy ion collision experiments. Such studies can also help us 
in understanding the internal structures of the dense astrophysical objects like neutron stars.

In order to perform the 
theoretical studies on the hadronic parameters at finite temperature, some non-perturbative approaches 
are required. One of the most applicable and powerful methods in this context is QCD sum rule approach. This method at zero 
temperature has been first proposed  in \cite{Shifman1,Shifman2}, which was 
based on the operator product expansion (OPE) and 
quark-hadron duality assumption. This approach has been extended to finite temperatures in \cite{Bochkarev}. At finite temperature, 
the Lorentz invariance is broken due to the residual O(3) symmetry. In order to restore the Lorentz invariance, 
the four-vector velocity of the medium is introduced. As a result, some new 
operators appear in the Wilson expansion \cite{Hatsuda,Mallik1,Shuryak} and the vacuum condensates are replaced by 
their thermal expectation values.

In the literature, there are a lot of studies on the investigation of the hadronic parameters at finite 
temperature, but only a few of them have been devoted to the analysis of the modifications of parameters such as 
the mass and residue of the nucleon at finite temperature \cite{Leutwyler,Adami,Kacir,Mallik2,Zakout,Abu-Shady1,Abu-Shady2,Abu-Shady3}. 
In \cite{Leutwyler}, 
the nucleon mass and its residue as a function 
of the temperature have been investigated using the correlation function of a three-quark current with the help of the virial 
expansion. It has been obtained that the nucleon mass rapidly rises with temperature, while its residue falls with 
increasing temperature. In \cite{Adami}, 
the nucleon mass at finite temperature has been analyzed via the 
QCD sum rules method. The authors have used Ioffe current and demonstrated that the nucleon mass 
is decrease with temperature. 
In \cite{Kacir}, the nucleon mass shift at finite temperature has been calculated using chiral counting arguments and a virial expansion. 
It has been shown that the nucleon's mass decreases with increasing the temperature up to roughly $T=90~MeV$, but after this point 
the mass starts to rapidly increase with increasing the temperature. 
In \cite{Mallik2}, the authors have obtained the spectral representation for the nucleon and given the expressions of the nucleon mass and its residue with 
respect to the temperature using Ioffe current by QCD sum rules approach. In this paper, S. Mallik and S. Sarkar have realized that the obtained results 
support the earlier results 
in \cite{Leutwyler}. In \cite{Zakout}, the nucleon mass at finite temperature has been calculated using Bethe-Salpeter equation (BSE) 
by observing that the nucleon mass decreases smoothly with temperature. In \cite{Abu-Shady1}, the nucleon properties below the critical point 
temperature have been investigated. The author has solved the field equations in the mean-field approximation using the effective mesonic 
potential at finite temperature and found that the nucleon mass increases up to $\frac{4}{5}T_{c}~MeV$ 
then decreases near to the critical temperature $T_{c}$. In \cite{Abu-Shady2}, the nucleon mass 
at nonzero temperature 
is investigated within the framework of the linear $\sigma$ model. It has been shown that the nucleon mass increases monotonically 
with the temperature $T$. In \cite{Abu-Shady3}, the nucleon properties at finite temperature have been analyzed in the linear 
sigma model, in which the higher-order mesonic interactions up to the eighth-order interactions are included. It has been 
found that the nucleon mass increases with increasing of temperature. 

In the present study, we calculate the mass and residue of the nucleon at finite temperature using the most general 
form of the interpolating 
current in the framework 
of thermal QCD sum rules. In our calculations, we use the thermal light quark propagator in coordinate space 
containing the perturbative and non-perturbative contributions as well as additional operators rising at finite temperature. 
We evaluate the two-point thermal correlation function first in coordinate space then transform the calculations to momentum 
space. 
In order to perform the 
numerical analysis, we firstly determine the working regions of the auxiliary parameters entering the calculations and use the 
fermionic and gluonic part of the energy density as well as  the temperature-dependent  light 
quark and gluon condensates together with the temperature-dependent  continuum threshold. We see that the Ioffe current which is a special case of the general current 
$(t=-1)$ remains roughly out of the reliable region. 

The outline of the paper is as follows: In the next section, taking into account the additional operators arising at finite 
temperature, we derive thermal QCD sum rules for the mass and residue of the 
nucleon. The last section is devoted to the numerical analyses of the sum rules and comparison of the results with those 
existing in the literature.

\section{Thermal QCD sum rules for the mass and residue of the nucleon}

This section is dedicated to the details of the calculations of the mass
and residue of the nucleon at finite temperature using thermal QCD sum
rules. The starting point is to consider the following two-point 
thermal correlation function:

\begin{eqnarray}\label{Eq1}
\Pi(p,T)=i \int d^{4}x~ e^{ip\cdot x} Tr\Big(\rho~ {\cal
T}\Big(J(x)\bar{J}(0)\Big)\Big),
\end{eqnarray}
where $p$ is the four momentum of the nucleon, $\rho=e^{-\beta H}/Tr
e^{-\beta H}$ is the thermal density matrix of QCD at temperature
$T=1/\beta$, $H$ is the QCD Hamiltonian and $\cal T$ indicates the time
ordering product. The most general form of the interpolating current for the nucleon can be written as
\begin{equation}
J(x)=2\epsilon_{abc}\sum_{i=1}^{2}\Bigg[q_{1}^{T,a}(x)CA_{1}^{i}q_{2}^{b}(x)
\Bigg]A_{2}^{i}q_{1}^{c}(x),
\end{equation}
where $a, b, c$ are color indices, $C$ is the charge conjugation
operator, $A_{1}^{1}=I$, $A_{1}^{2}=A_{2}^{1}=\gamma_5$ and 
$A_{2}^{2}=t$, with $t$ is being  an arbitrary auxiliary parameter
and $t=-1$ corresponds to the Ioffe current (see \cite{thomas,drukarev2013,leinweber,stein} for detailed discussions on the nucleon 
current). Considering the isospin symmetry, we will deal with the proton ($q_1=u$ and $q_2=d$) in the following. 

According to the general philosophy of the QCD sum rule
approach, the above mentioned thermal correlation function can be calculated in two
different ways. First, in terms of the hadronic parameters called the hadronic side and the second, in terms of QCD degrees of 
freedom called the OPE side. 
Equating the hadronic and OPE sides of the correlation
function, the sum rules for the mass and residue of the nucleon at
finite temperature are obtained. Applying a Borel transformation and continuum subtraction to both sides of 
the obtained sum rules, the contributions of the
higher states and continuum are suppressed.

In order to obtain the hadronic side of the
correlation function, we insert a complete set of the nucleon state with
the same quantum numbers as the interpolating current into the thermal
correlation function. This side can be written in terms of different structures as  
\begin{eqnarray}\label{hadstruction}
\Pi^{Had}(p,T)&=&\Pi^{Had}_{p}(p^2,p_0,T)\!\not\!{p}+\Pi^{Had}_{u}(p^2,p_0,T)\!\not\!{u}\nonumber \\
&+&\Pi^{Had}_{p,u}(p^2,p_0,T)\!\not\!{p}\!\not\!{u}+\Pi^{Had}_{S}(p^2,p_0,T)I,
\end{eqnarray}
where $u_{\mu}$ is the four-velocity of the heat bath, $p_0$ is the energy of the quasi-particle and $I$ is the unit
matrix. In the rest frame of the heat bath, $u_{\mu}=(1,0,0,0)$
and $u^2=1$. To obtain the mass and residue of the nucleon, we use the structure $\not\!p$ whose coefficient is 
\begin{eqnarray}\label{correpslash}
\Pi^{Had}_p(p^2,p_0,T)&=&-\lambda_N^{2}(T)\frac{1}{p^2-m_{N}^{2}(T)},
\end{eqnarray}
where $m_{N}(T)$ is the nucleon mass and $\lambda_{N}(T)$ is its residue at finite temperature. After performing the Borel 
transformation with respect to $p^2$ in Eq. (\ref{correpslash}), the final form of the coefficient of $\not\!p$ on 
hadronic side is obtained as  
\begin{eqnarray}
\hat{B}\Pi^{Had}_p(p_0,T)=-\lambda_N^{2}(T)e^{-m_{N}^{2}(T)/M^2},
\end{eqnarray}
where $M^2$ is the Borel mass parameter.

On the other hand, the OPE side of the thermal
correlation function is calculated via OPE  in deep Euclidean region.  The correlation function on this side can also be written 
in terms of different structures as 
\begin{eqnarray} \Pi^{OPE}(p,T)&=&\Pi_{p}^{OPE}(p^2,p_0,T)\!\not\!{p}+\Pi_{u}^{OPE}(p^2,p_0,T)\!\not\!{u} \nonumber \\
&+&\Pi_{p,u}^{OPE}(p^2,p_0,T)\!\not\! {p}\!\not\! {u}+\Pi_{S}^{OPE}(p^2,p_0,T)I.
\end{eqnarray}
To calculate the OPE side, we use the explicit form of the interpolating current in the thermal correlation function and contract 
out all quark pairs using the Wick's theorem. As a result, we get
\begin{eqnarray}\label{corre2}
\Pi^{OPE}(p,T) &=& 4i\epsilon_{abc}\epsilon_{a'b'c'}\int d^4 x
e^{ipx}\Bigg\langle 
\Bigg\{\Bigg(\gamma_{5}S^{cb'}_{u}(x)S'^{ba'}_{d}(x)S^{ac'}_{u}(x)\gamma_{5}
\nonumber \\
&-&\gamma_{5}S^{cc'}_{u}(x)\gamma_{5}Tr\Bigg[S^{ab'}_{u}(x)S'^{ba'}_{d}(x)\Bigg]
\Bigg)
+t\Bigg(\gamma_{5}S^{cb'}_{u}(x)\gamma_{5}S'^{ba'}_{d}(x)S^{ac'}_{u}
(x)\nonumber \\
&+&S^{cb'}_{u}(x)S'^{ba'}_{d}(x)\gamma_{5}S^{ac'}_{u}(x)\gamma_{5}
-\gamma_{5}S^{cc'}_{u}(x)Tr\Bigg[S^{ab'}_{u}(x)\gamma_{5}S'^{ba'}_{d}(x)\Bigg]
\nonumber \\
&-&S^{cc'}_{u}(x)\gamma_{5}Tr\Bigg[S^{ab'}_{u}(x)S'^{ba'}_{d}(x)\gamma_{5}\Bigg]
\Bigg)
+t^2\Bigg(S^{cb'}_{u}(x)\gamma_{5}S'^{ba'}_{d}(x)\gamma_{5}S^{ac'}_{u}
(x)\nonumber \\
&-&S^{cc'}_{u}(x)Tr\Bigg[S^{ba'}_{d}(x)\gamma_{5}S'^{ab'}_{u}(x)\gamma_{5}\Bigg]
\Bigg) \Bigg\} \Bigg\rangle_{T},
\end{eqnarray}
where $S'=CS^TC$, $S_{q}(x)$ is the thermal light quark propagator and
$Tr[...]$ stands for the trace of gamma matrices. The
thermal light quark propagator is given as 
\begin{eqnarray}\label{lightquarkpropagator}
S_{q}^{ij}(x)&=& i\frac{\!\not\!{x}}{ 2\pi^2 x^4}\delta_{ij}
-\frac{m_q}{4\pi^2 x^2}\delta_{ij}-\frac{\langle
\bar{q}q\rangle}{12}\delta_{ij} -\frac{ x^{2}}{192} m_{0}^{2}
\langle
\bar{q}q\rangle\Big[1-i\frac{m_q}{6}\!\not\!{x}\Big]\delta_{ij}
\nonumber\\
&+&\frac{i}{3}\Big[\!\not\!{x}\Big(\frac{m_q}{16}\langle
\bar{q}q\rangle-\frac{1}{12}\langle u\Theta^{f}u\rangle\Big)
+\frac{1}{3}\Big(u\cdot x\!\not\!{u}\langle
u\Theta^{f}u\rangle\Big)\Big]\delta_{ij}
\nonumber\\
&-&\frac{ig_s \lambda_{A}^{ij}}{32\pi^{2} x^{2}}
G_{\mu\nu}^{A}\Big(\!\not\!{x}\sigma^{\mu\nu}+\sigma^{\mu\nu}
\!\not\!{x}\Big),
\end{eqnarray}
where $m_{q}$ is the light quark mass, $\langle\bar{q}q\rangle$ is the
temperature-dependent quark condensate, $G_{\mu\nu}^{A}$ is the external gluon field at finite temperature, 
$\Theta^{f}_{\mu\nu}$ is 
the fermionic part of the energy momentum tensor and $\lambda_{A}^{ij}$
are Gell-Mann matrices with $\lambda_{A}^{ij}\lambda_{A}^{kl}=\frac{1}{2}(\delta_{il}\delta_{jk}-\frac{1}{N_{c}}\delta_{ij}\delta_{kl})$. 
We insert the above quark propagator into Eq. (\ref{corre2}) and perform
the four-integral over $x$ to go to the momentum space using Fourier
transformations.  To suppress the contribution of the higher states
and continuum, a Borel transformation with respect to the $p^{2}$
as well as continuum subtraction are applied. We also use 
\begin{eqnarray}\label{TrGG} 
\langle Tr^c G_{\alpha \beta} G_{\mu \nu}\rangle &=& \frac{1}{24} (g_{\alpha \mu} g_{\beta \nu} -g_{\alpha
\nu} g_{\beta \mu})\langle G^a_{\lambda \sigma} G^{a \lambda \sigma}\rangle \nonumber \\
 &+&\frac{1}{6}\Big[g_{\alpha \mu}
g_{\beta \nu} -g_{\alpha \nu} g_{\beta \mu}-2(u_{\alpha} u_{\mu}
g_{\beta \nu} -u_{\alpha} u_{\nu} g_{\beta \mu} -u_{\beta} u_{\mu}
g_{\alpha \nu} +u_{\beta} u_{\nu} g_{\alpha \mu})\Big]\nonumber \\
&\times&\langle u^{\lambda} {\Theta}^g _{\lambda \sigma} u^{\sigma}\rangle 
\end{eqnarray}
to relate the two-gluon condensate to the gluonic part of the energy momentum tensor.
Here $\Theta^{g}_{\lambda \sigma}$ is the traceless gluonic part
of the stress-tensor of the QCD. After lengthy calculations, the $\Pi_{p}^{OPE}$ function in Borel scheme up to 
linear terms in light quark mass is obtained as 
\begin{eqnarray}\label{borelpi}
\hat{B}\Pi_{p}^{OPE}(p_0,T)&=&\frac{5 + 2t+ 5t^2}{512\pi^4}\int^{s_0(T)}_{(2m_{u}+m_{d})^2}ds \exp\Big(-\frac{s}{M^2}\Big)s^2\nonumber\\
&+&\frac{\langle\bar{q}q\rangle}{96\pi^2}\Bigg[m_{0}^2\Big((-23 - 2t+ 13t^2)m_{d} + 2(-11 - 8t + 7t^2)m_{u}\Big)\nonumber\\
&+&\Big((51 + 6t - 21t^2)m_{d}- 18(-3 - 2t +t^2)m_{u}\Big)\int^{s_0(T)}_{(2m_{u}+m_{d})^2}ds\exp\Big(-\frac{s}{M^2}\Big)\Bigg]\nonumber\\
&+&\frac{\langle u\Theta^{f} u \rangle}{72\pi^2}(5 + 2t+ 5t^2)\Bigg[8p_{0}^2-5\int^{s_0(T)}_{(2m_{u}+m_{d})^2}ds\exp\Big(-\frac{s}{M^2}\Big)\Bigg]\nonumber\\
&-&\frac{\alpha_{s} \langle u\Theta^{g} u \rangle}{192\pi^3 N_{c}}(5 + 2t + 5t^2)(-1 +N_{c})\Bigg[4p_{0}^2 -\int^{s_0(T)}_{(2m_{u}+m_{d})^2}ds\exp\Big(-\frac{s}{M^2}\Big)\Bigg]\nonumber\\
&-&\frac{3\langle\alpha_{s}G^2\rangle}{256\pi^3 N_{c}}(5 + 2t+ 5t^2)(-1 + N_{c})\int^{s_0(T)}_{(2m_{u}+m_{d})^2}ds\exp\Big(-\frac{s}{M^2}\Big)\nonumber\\
&-&\langle\bar{q}q\rangle^2\Bigg[-\frac{(-5 - 2t + 7t^2)(m_{0}^2 - 2M^2)}{12M^2}\Bigg]\nonumber \\
&+&\frac{\langle\bar{q}q\rangle\langle u\Theta^{f} u \rangle}{27M^6}\Bigg\{-2m_{0}^2\Big((1 +t^2)m_{d}+2(2 + t+ 2t^2)m_{u}\Big)(M^2 + 2p_{0}^2)\nonumber\\
&+&3M^2\Bigg[m_{d}\Big((-7 - 2t + 5t^2)M^2 + 8(1 +t^2)p_{0}^2\Big)\nonumber\\
&+&2m_{u}\Big(M^2 + 7t^2M^2 + 8(2 + t + 2t^2)p_{0}^2\Big)\Bigg]\Bigg\}\nonumber\\
&+&\frac{\alpha_{s}\langle\bar{q}q\rangle\langle u\Theta^{g} u \rangle}{144\pi N_{c}M^6}\Bigg[\Big((1 + t)^2m_{d}- 2(3 + 2t + 3t^2)m_{u}\Big)(-1 + N_{c})\nonumber\\
&\times&\Big(m_{0}^2(3M^2-2p_{0}^2)+ 3M^2(-5M^2+4p_{0}^2)\Big)\Bigg]\nonumber\\
&+&\frac{\langle\bar{q}q\rangle\langle\alpha_{s}G^2\rangle}{192\pi N_{c}M^4}\Big[(1 + t)^2m_{d}- 2(3 + 2t + 3t^2)m_{u}\Big]
(4m_{0}^2-15M^2)(-1 + N_{c}) \nonumber\\
&-&\frac{\alpha_{s} \langle u\Theta^{f} u \rangle \langle u\Theta^{g} u \rangle}{36\pi N_{c}M^4}(-1 + N_{c})\Big[(21 + 2t+ 21t^2)M^2 - 4(7 + 6t+ 7t^2)p_{0}^2\Big]\nonumber\\
&-&\frac{\langle\alpha_{s}G^2\rangle \langle u\Theta^{f} u \rangle}{144\pi N_{c}M^4}(-1 + N_{c})\Big[(71 + 22t + 71t^2)M^2 - 8(1 + t)^2p_{0}^2\Big]\nonumber\\
&+&\frac{2\langle u\Theta^{f} u \rangle^2}{9 M^2}(5 + 2t + 5t^2)
\end{eqnarray}
Finally, matching the hadronic and OPE sides of the 
correlation function, the sum rule for the residue of the nucleon
at finite temperature is obtained as:
\begin{eqnarray}\label{residuesumrule}
-\lambda_N^{2}(T)e^{-m_{N}^2(T)/M^2}=\hat{B}\Pi_{p}^{OPE}(p_0,T).
\end{eqnarray}
To find the mass of nucleon from above sum rule, we apply
derivative with respect to $1/M^2$ to both sides of Eq.
(\ref{residuesumrule}) and divide by itself. Eventually, the mass of
nucleon at finite temperature is obtained as:
\begin{eqnarray} \label{ratio}
m_{N}^2(T)=\frac{\frac{\partial}{\partial(-\frac{1}{M^2})}\Big[\hat{B}\Pi_{p}^{OPE}(p_0,T)\Big]}{\hat{B}\Pi_{p}^{OPE}(p_0,T)}.
\end{eqnarray}

\section{Numerical results}%

In this section, we present numerical analysis on the mass and
residue of the nucleon at finite temperature obtained via thermal QCD
sum rules. We determine how the results
at finite temperature shift from those obtained at zero temperature. For this aim, we use the input parameters given in Table 1.

\begin{table}[ht!]
\centering
\rowcolors{1}{lightgray}{white}
\begin{tabular}{cc}
\hline \hline
   Parameters  &  Values    
           \\
\hline \hline
$p_0   $          &  $1~GeV$     \\
$ m_{u}   $          &  $(2.3_{-0.5}^{+0.7})$ $MeV$      \\
$ m_{d}   $          &   $(4.8_{-0.3}^{+0.5})$ $MeV$    \\
$ m_{0}^{2}   $          &  $(0.8\pm0.2)$ $GeV^2$         \\
$ \langle0|\overline{q}q|0\rangle_{q=u,d} $          &  $-(0.24\pm0.01)^3$ $GeV^3$       \\
$ {\langle}0\mid \frac{1}{\pi}\alpha_s G^2 \mid 0{\rangle}$          &  $ (0.012\pm0.004)~GeV^4$   \\
\hline \hline
\end{tabular}
\caption{The values of input parameters used in numerical calculations \cite{Olive,H.G.Dosch1,H.G.Dosch2,B.L.Ioffe}. }
\end{table}
Besides these input parameters, we need to know 
the temperature dependence of the quark and gluon condensates, the thermal average of the energy density 
and the temperature-dependent expression of 
continuum threshold.

 In this study, for the  temperature-dependent quark condensate we use the parametrization given in \cite{ayala1}, which can be fitted to the following function
up to the  quark-gluon deconfinement at a critical temperature $T_c\simeq 197~MeV$:
\begin{eqnarray}\label{qbarq}
\langle\bar{q}q\rangle=
\frac{\langle0|\bar{q}q|0\rangle}{1+e^{18.10042 (1.84692 T^2+4.99216 T-1)}},
\end{eqnarray}
where $\langle0|\bar{q}q|0\rangle$ denotes the  light-quark condensate in vacuum. The above parametrization is consistent with the lattice QCD results \cite{Bazavov,cheng}.

%
%

For the gluonic and fermionic  parts of the energy density we use the results obtained from
lattice QCD at higher temperatures. In the rest frame of the heat bath,
the results of some quantities obtained using lattice QCD in \cite{M.Cheng} are well fitted by the help
of the following parametrization:
\begin{eqnarray}
 \langle \Theta^g_{00}\rangle=  \langle \Theta^f_{00}\rangle=T^4e^{(113.867T^2-12.190T)}-10.141T^5,
\end{eqnarray}
where this is valid for $T\geq130~MeV $.

%
%
%
%

We take the temperature-dependent continuum threshold 
as 
%
\begin{equation}\label{eqn16}
s_{0}(T)\simeq s_{0}\frac{\langle\bar{q}q\rangle}{\langle0|\bar{q}q|0\rangle},  \\
\end{equation}
where $s_{0}$ is the continuum threshold at zero temperature. The
temperature-dependent gluon condensate corresponding to $T_c\simeq 197~MeV$ is also given  as \cite{ayala2}
%
\begin{eqnarray}\label{G2TLattice}
\langle G^2\rangle=\langle
0|G^2|0\rangle\Big[1-1.65(\frac{T}{T_c})^{8.735}+0.04967(\frac{T}{T_c})^{0.7211}\Big],
\end{eqnarray}
where $\langle0|G^2|0\rangle$ stands for the gluon condensate in vacuum.
%
%
%

The next step is to determine the working regions for three auxiliary parameters, namely 
Borel mass parameter $M^{2}$, the continuum threshold $s_{0}$ and mixing parameter $t$. It is expected that 
the physical quantities are roughly independent of these auxiliary parameters in their working regions. 

The working region for the Borel mass 
parameter is found demanding that the contributions of the higher states 
and continuum are adequately suppressed as well as the contributions of the higher 
dimensional operators are small. These conditions lead to the 
interval $0.8~GeV^2\leqslant M^2 \leqslant 1.2~GeV^2$ for the Borel mass parameter. The continuum threshold is not utterly 
arbitrary but it is associated with the energy of the first excited state with the same
quantum numbers as the chosen interpolating current. From our numerical analysis, we obtain the interval 
$s_{0}=(1.5-2.0)~ GeV^2$ for the continuum threshold. In order to determine the working region of the mixing parameter 
$t$, we need to investigate the dependence of the physical quantities on $t$ in the 
whole range $-\infty\leq t\leq\infty$. For this purpose, we plot the mass and residue of the nucleon with 
respect to $x=cos\theta$, 
where $t=tan\theta$, at fixed values of the continuum threshold $s_0$ for Borel mass parameter $M^{2}=1~GeV^2$ at $T=0$ in figure 1. 
From this figure, we see that the dependences of the mass and residue on this parameter are relatively weak in the 
intervals $-0.6\leq x \leq-0.2$ and $0.2\leq x \leq0.6$. Also, we realize that the Ioffe current which corresponds to $x=-0.71$ 
remains out of the reliable region.

 \begin{figure}[h!]
\begin{center}
\includegraphics[width=8cm]{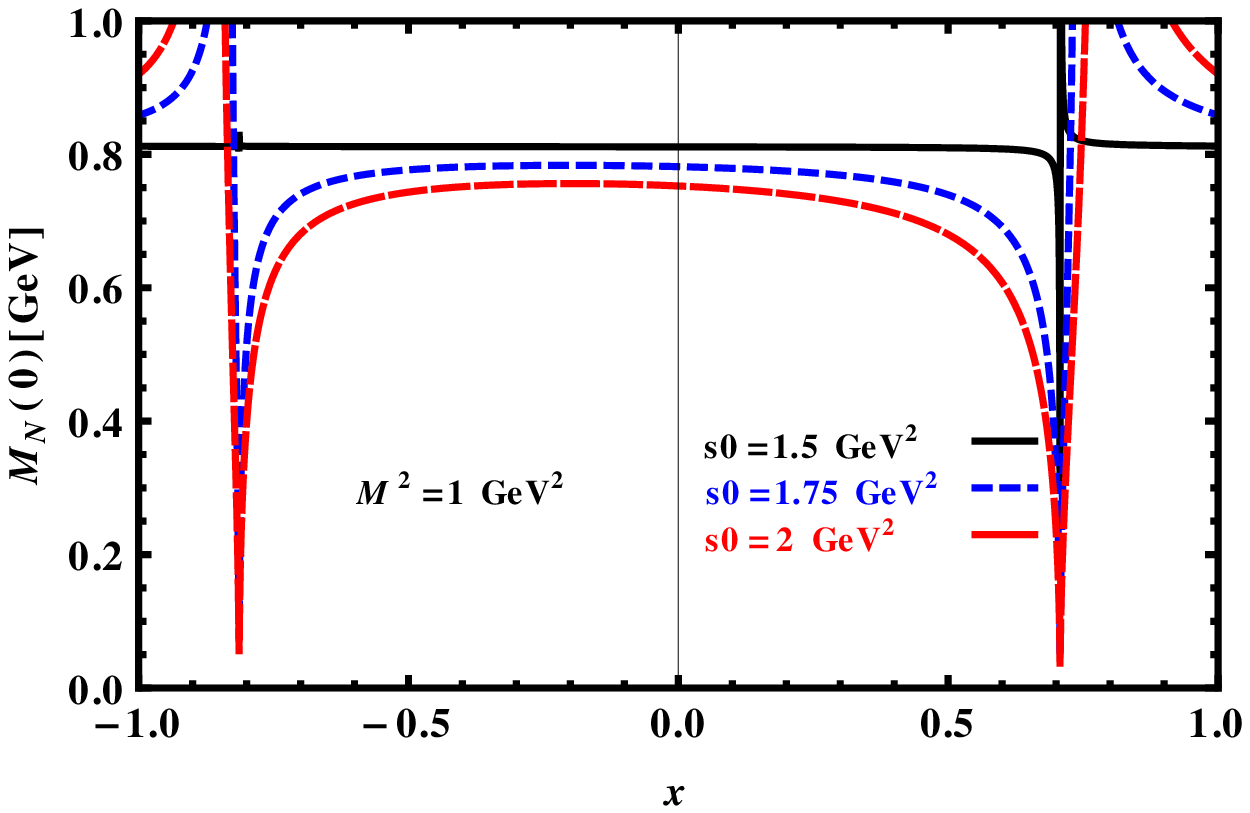}
\includegraphics[width=8cm]{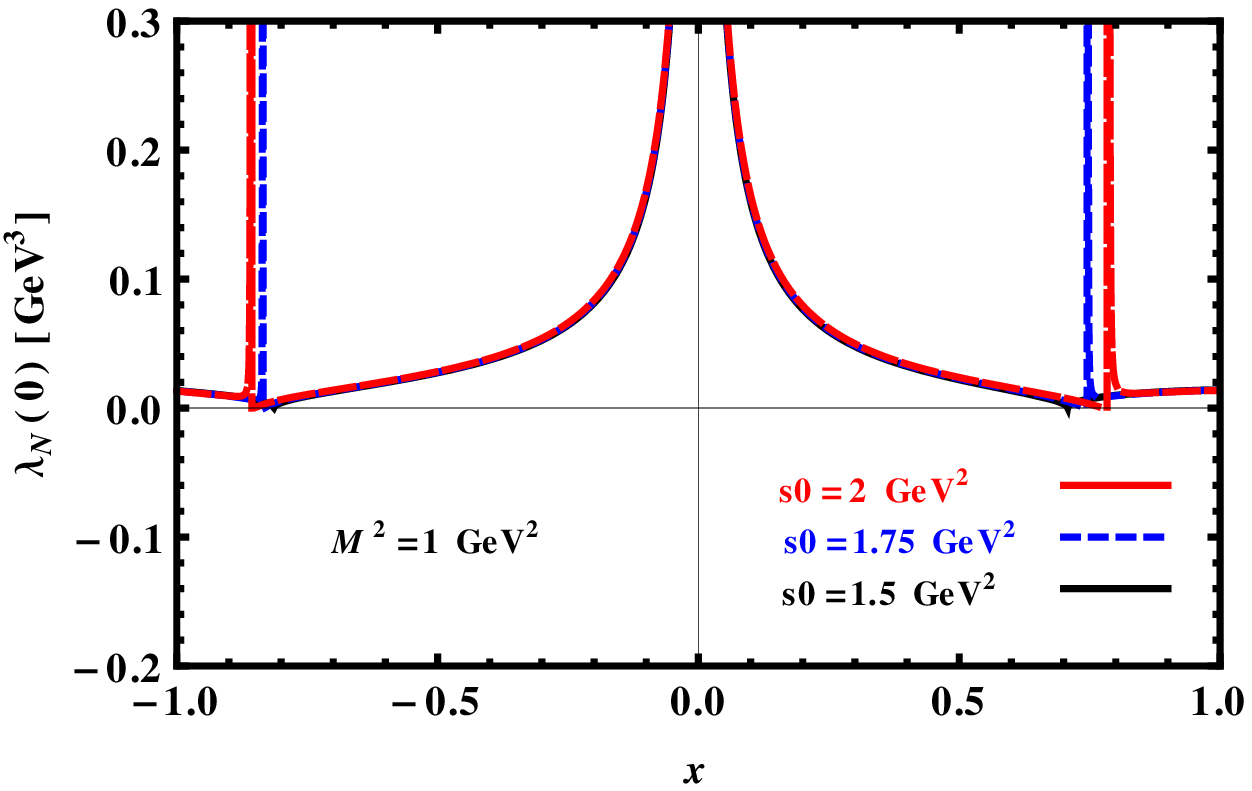}
\end{center}
\caption{The mass and residue of the nucleon as a function of $x$ at $M^2=1~GeV^2$, $T=0$ and at different fixed values of 
$s_{0}$.} \label{Diagrams1}
\end{figure}

To see how our results depend on the Borel mass parameter in the above determined working 
regions, we plot 
the mass and residue of the nucleon at zero temperature with respect to Borel mass parameter $M^{2}$ at different fixed values
of $x$ and $s_0$ in figure 2. This figure depicts that the values of the physical quantities under consideration 
depend weakly on the 
Borel parameters in its working region, $0.8~GeV^2\leqslant M^2 \leqslant 1.2~GeV^2$. 

 \begin{figure}[h!]
\begin{center}
\includegraphics[width=8cm]{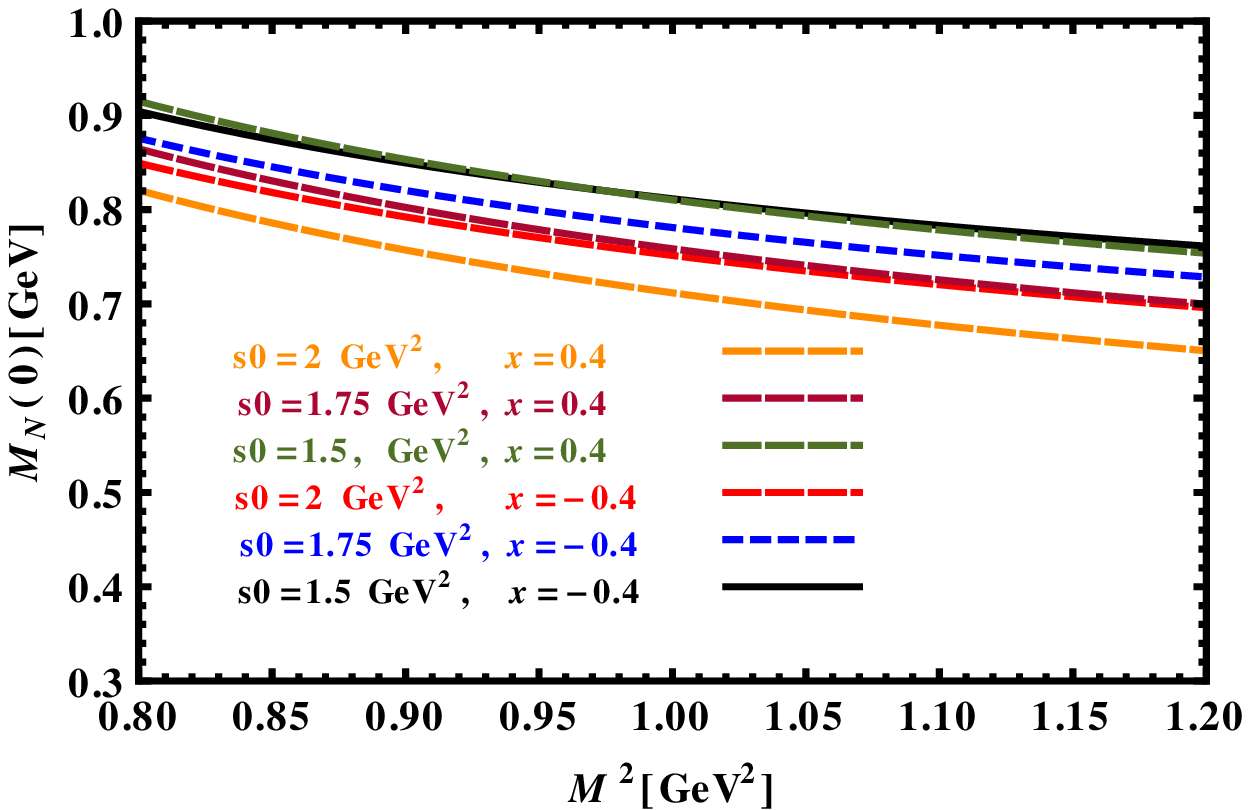}
\includegraphics[width=8cm]{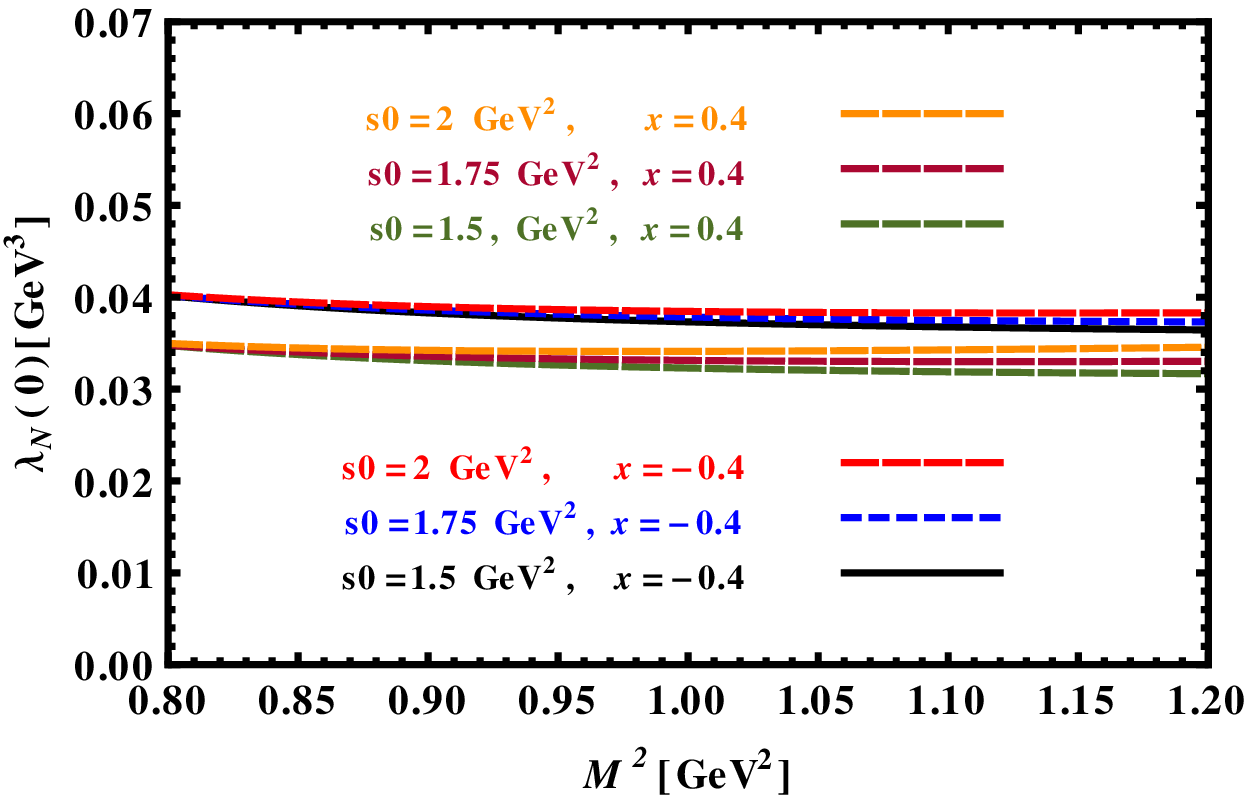}
\end{center}
\caption{The mass and residue of the nucleon as a function of $M^2$ at $T=0$ and at different fixed values of 
$s_{0}$ and $x$.} \label{Diagrams1}
\end{figure}

 \begin{figure}[h!]
\begin{center}
\includegraphics[width=8cm]{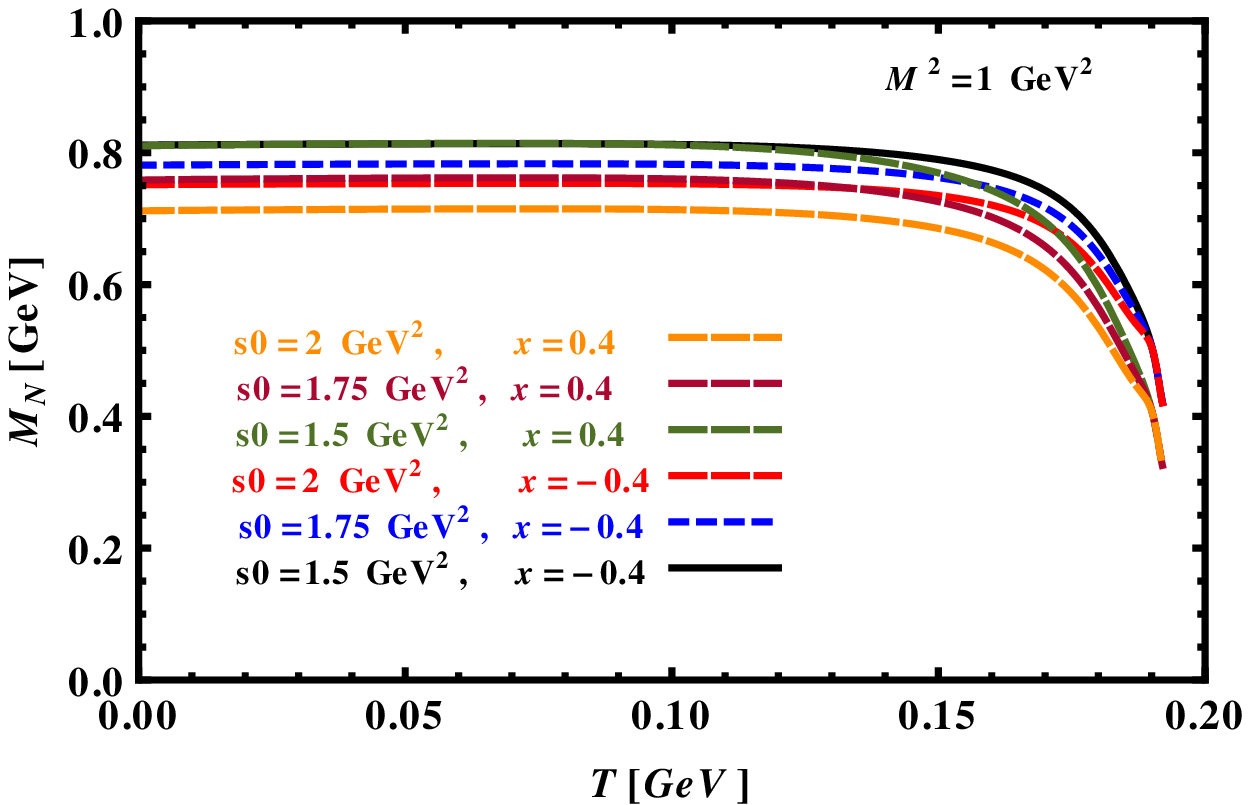}
\includegraphics[width=8cm]{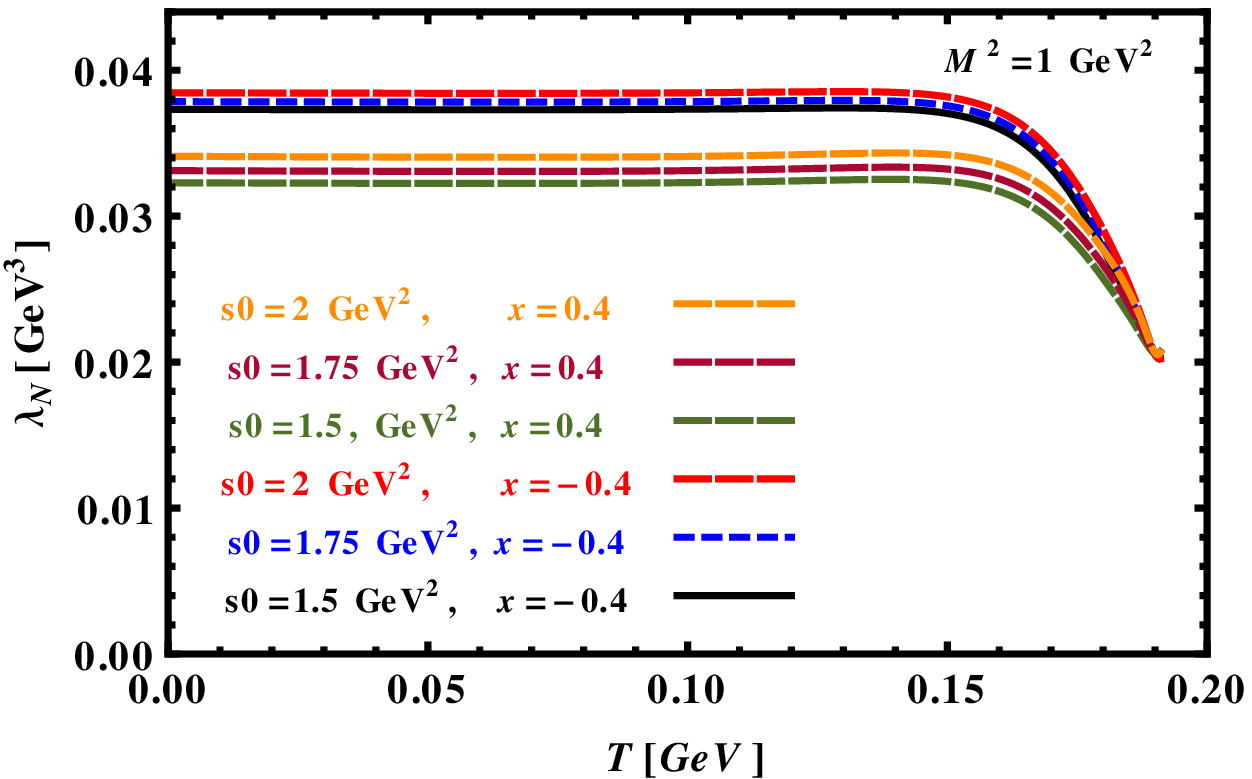}
\end{center}
\caption{The mass and residue of the nucleon as a function of temperature at $M^{2}=1~GeV^2$ and 
at different fixed values of $s_{0}$ and $x$.
} \label{Diagrams1}
\end{figure}

\begin{figure}[h!]
\begin{center}
\includegraphics[width=8cm]{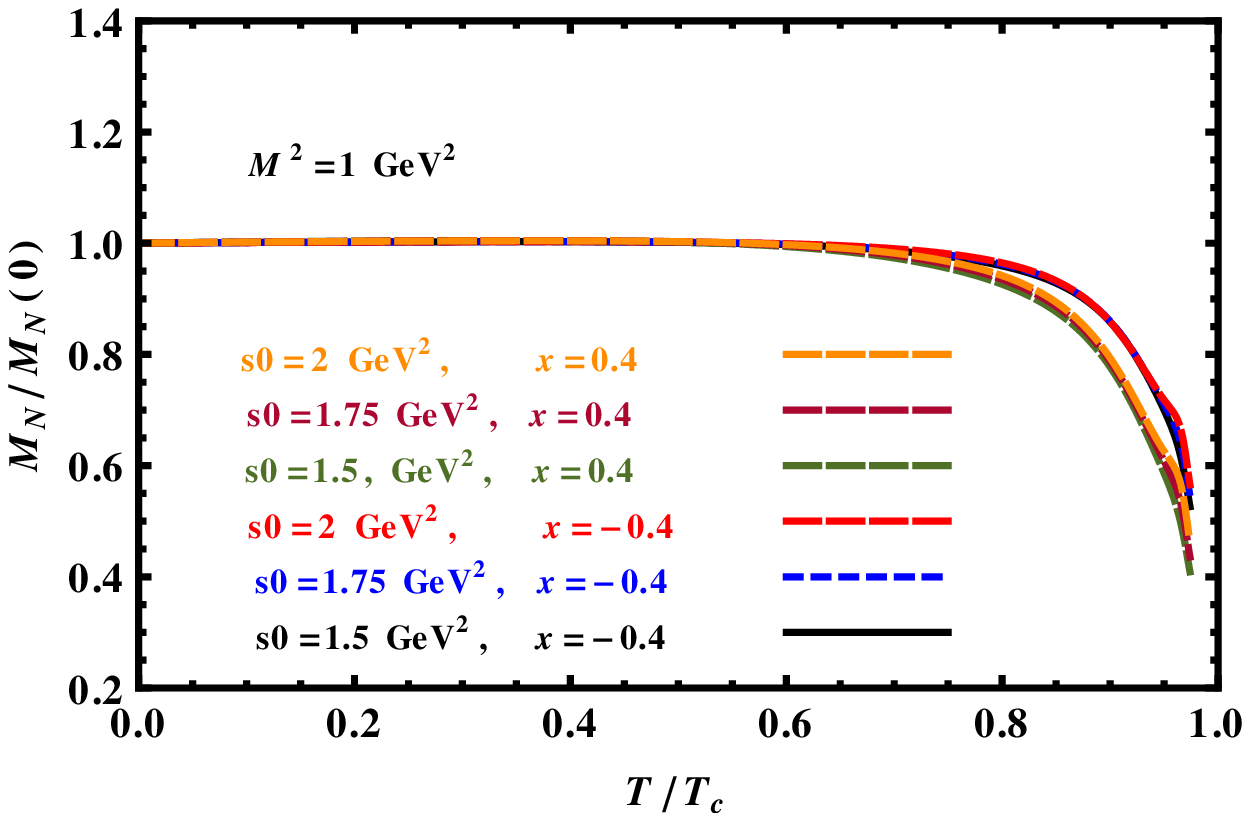}
\includegraphics[width=8cm]{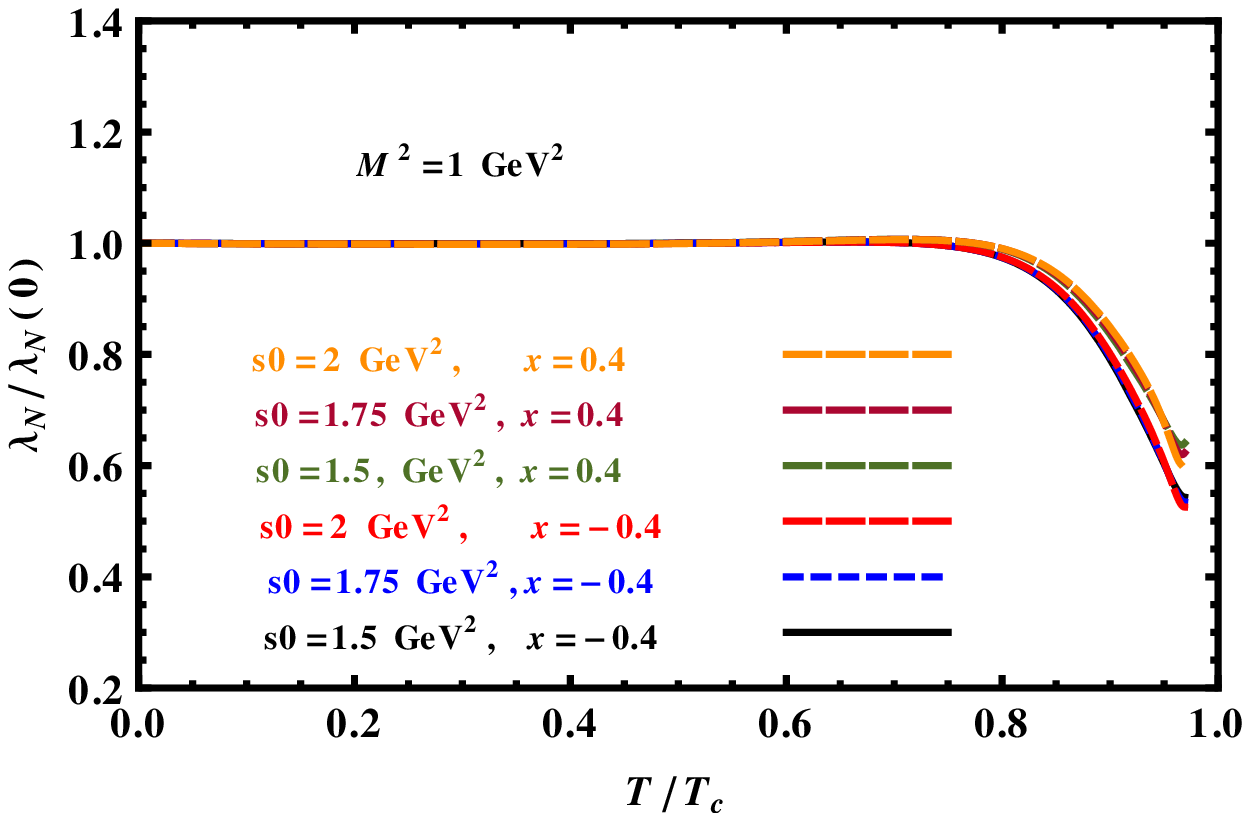}
\end{center}
\caption {$M_{N}/M_{N}(0)$ and $\lambda_{N}/\lambda_{N}(0)$ as a function of $T/T_c$ at $M^2=1$ $GeV$ and 
at different fixed values of $s_{0}$ and $x$.
} \label{Diagrams1}
\end{figure}

In further analysis, the dependence of the mass and residue of the nucleon on the temperature are shown in figure 3 at different 
values of the 
auxiliary parameters in their working regions. With a quick glance at this figure, we observe 
that the mass and residue of the nucleon remain approximately unchanged up to $T\cong0.15~ GeV$, after which they  start to  diminish increasing the temperature. 
Near to the critical temperature, the mass of the nucleon falls to roughly $42\%$ of its values at zero temperature while the residue reaches roughly to $60\%$ of its value at zero temperature. 

To better see the deviations of the mass and residue of the nucleon from their vacuum values, we show the dependence of the 
ratios $M_{N}/M_{N}(0)$ and $\lambda_{N}/\lambda_{N}(0)$ to $T/T_c$ in figure 4. When we compare these behaviors in 
terms of the temperature with those of existing results in the literature, we see that as far as the residue is concerned our 
results support the behaviors obtained in \cite{Leutwyler,Mallik2} for the residue with respect to the temperature at whole region. In the case of mass, our 
results on the variation of the mass of the nucleon are in agreement with those of \cite{Adami,Zakout}. Although are also in good agreements at low temperatures, our results on the   behavior of the mass with respect to temperature
 are in contradiction with  those of 
\cite{Leutwyler,Kacir,Mallik2,Abu-Shady1,Abu-Shady2,Abu-Shady3}  at high temperatures. 

To summarize, we calculated the mass and residue of the nucleon at finite temperature in the framework of the two-point 
thermal QCD sum rules. We used the general form of the nucleon's interpolating current and found the working regions of 
auxiliary parameters entering the calculations. It has been obtained that the Ioffe current remains out of the reliable region for the 
general mixing parameter entering the general form of the interpolating current. We have also used the temperature 
dependence of the quark and gluon condensates, the thermal average of the energy density 
and the temperature-dependent expression of continuum threshold to numerically analyze the sum rules for the mass and 
residue of the nucleon. We found that the mass and residue of the nucleon remain unchanged up to roughly $0.15~ GeV$. 
After this point, they start to decrease with increasing the temperature such that near to the critical 
temperature they  fall down, considerably. This can be consider as a signal for transition to the quark-gluon plasma phase. We also 
compared the behavior of the mass and residue of the nucleon in terms of temperature obtained in the present work with those existing in the 
literature.

\section{Acknowledgment}
This work has been supported in part by the Scientific and Technological
Research Council of Turkey (TUBITAK) under the National Postdoctoral Research Scholarship Programme 2218.

\end{document}